\documentclass[useAMS,usenatbib]{mn2e}

\usepackage[pdftex]{graphicx}
\usepackage[utf8]{inputenc}
\usepackage{amsmath}
\usepackage{amssymb}
\usepackage{color}
\usepackage{tabularx}
\usepackage{sidecap}

\newcommand{\nbody}{\mbox{N{\sc body} 6} }
\newcommand{\Msun}{\text{M}_{\sun}}

\renewcommand{\d}{\text{d}}
\newcommand{\mnras}{MNRAS}
\newcommand{\apj}{ApJ}
\newcommand{\comment}[1]{}

\title[The evolution of the surface brightness]{The evolution of the surface brightness of a star cluster as a result of residual star-forming gas expulsion}
\author[F. L\"ughausen et al.]
{F.~L\"ughausen,$^1$\thanks{fabian@astro.uni-bonn.de} 
G.~Parmentier,$^{1,2}$\thanks{gparm@mpifr-bonn.mpg.de}
J.~Pflamm-Altenburg,$^1$\thanks{jpflamm@astro.uni-bonn.de} 
and P.~Kroupa$^1$\thanks{pavel@astro.uni-bonn.de}
\\
$^1$Argelander-Institut f\"ur Astronomie, Auf dem H\"ugel 71, D-53121 Bonn, Germany\\
$^2$Max-Planck-Institut f\"ur Radioastronomie, Auf dem H\"ugel 69, D-53121 Bonn, Germany}

\begin{document}

  \date{\today}
  \maketitle

  \begin{abstract}
    Direct $N$-body calculations are presented of the early evolution of exposed clusters to quantify the influence of gas expulsion on the time-varying surface brightness. By assuming that the embedded OB stars drive out most of the gas after a given time delay, the change of the surface brightness of expanding star clusters is studied. The influence of stellar dynamics and stellar evolution is discussed.\\
    The growth of the core radii of such models shows a remarkable core re-virialisation. The decrease of the surface mass density during gas expulsion is large and is only truncated by this re-virialisation process. However, the surface brightness within a certain radius does not increase noticeably. Thus, an embedded star cluster cannot reappear in observational surveys after re-virialisation. This finding has a bearing on the observed infant mortality fraction.
  \end{abstract}



    \section{Introduction}
Since stars form in dense clumps within molecular clouds, very young clusters are observed to be embedded in gas. Typically $1/3$ of the initial cluster-forming gas mass is converted into stars \citep{Lada:2003il,Machida:2011uw}. Following star formation, and especially after formation of OB stars, most of the residual gas is driven out, so that there is nearly no gas observed in populations older than a few million years. While the gas ejection process begins, self-gravitation decreases, resulting in a rapid expansion of the young star cluster (\citealt{1997MNRAS.284..785G}, \citealt{2006MNRAS.373..752G}, \citealt{2003MNRAS.338..665B}, \citealt{2007MNRAS.380.1589B}).
In this contribution we investigate cluster formation and evolution during this process, focusing on the surface brightness of star clusters responding to residual star-forming gas expulsion. We want to answer the question of how the surface brightness is influenced by a) stellar dynamics and b) stellar evolution, and how both relate to each other.

In the context of gas expulsion, numerical $N$-body computations show a re-virialisation process taking place after a few million years in the centre of the system. While the cluster outskirts continue to expand, the centre contracts \citep*{2001MNRAS.321..699K}, leading to an increase of the density in the central region. Since at any given age, the surface mass density has an impact on the surface brightness, this process is of special interest. We ask what role this effect plays in the evolution of the surface brightness and whether its impact is sufficient to let the cluster core disappear under a given detection limit owing to its expansion, then re-appear after re-virialisation.

We have performed direct $N$-body computations, including stellar evolution, with initial conditions similar to \cite{2001MNRAS.321..699K}, with the emphasis being put on the time-evolution of the cluster surface brightness and surface mass density. The residual gas is modelled by a time-dependent background potential.


    \section{The Code}
		The main aspect here is the impact of gas expulsion on the early evolution of star clusters. We choose \nbody \cite[]{2000chun.proc..286A} to perform these direct $N$-body simulations since this code already offers features that enable us to account for gas expulsion. In doing so, a time-varying analytical background potential is added which represents the gas and therefore its gravitational interaction with the stellar population. We have to make this assumption to simplify the complex hydrodynamical processes in the gas component. However, it has been concluded by \citet{2001ASPC..230..311G} that this approximation is physically realistic.
		
		In addition to the gas potential, a standard solar-neighbourhood tidal field is adopted. 
		Also, stellar evolution is taken into account, because it is important when determining the luminosity of stars.
		We did not include pre-main sequence evolution as it is not part of the \nbody package yet, and incorporating it into \nbody is a very major effort. This means that the star clusters would initially be brighter than we compute, but also that they would fade even more strongly. The brightening of some of the clusters listed in Tab. \ref{tab:brightness_changes_revirialisation} are therefore upper limits, since pre-main sequence evolution would reduce this brightening because the stars fade towards the main sequence. The brightening effect due to revirialisation is therefore even less observable than estimated by our present models such that clusters cannot reappear after revirialisation.

\subsection{Analytical gas model}\label{gas_model}
	We assume that both the stars and gas obey a Plummer density profile \citep{1911MNRAS..71..460P} with equal half-mass radii. 
The star formation efficiency (SFE) at the onset of gas expulsion is $\text{SFE} = 1/3$, that is, the gas mass is twice the initial stellar mass $M_\text{st}(t<t_\text{D})$, 
\begin{equation}
	M_\text{g}(t<t_\text{D}) = 2\, M_\text{st}(t<t_\text{D}) \, .
\end{equation}

A star with a position $\textbf r$ relative to the centre of the cluster experiences an acceleration from the background potential, 
\begin{equation}\label{eq:acceleration}
 \textbf a_\text{g} = - \frac{GM_\text{g}(t)}{(r^2+r_\text{pl,g}^2)^{3/2}}\, \textbf r,
\end{equation}
where $M_\text{g}(t)$ is the time-varying mass and $r_\text{pl,g}$ is the Plummer radius of the cluster-forming region. $G$ is the gravitational constant. 

Gas expulsion is modelled by decreasing the gas mass, the Plummer radius $r_\text{pl,g}$ remaining unchanged. The gas mass $M_\text{g}(t)$ -- and accordingly the background potential -- is assumed to stay constant until a time-delay $t_\text{D} = 0.6\,\text{Myr}$.
After this time delay it decreases following
\begin{equation}
 M_\text{g}(t) = \frac{M_\text{g}(t<t_\text{D})} {1+\left(t-t_\text{D}\right)/\tau },\label{eq:gas-mass-vs-time}
\end{equation}
where $\tau$ is the gas expulsion time scale, which depends on the velocity of the heated ionized gas \citep[$10\,\text{km}\,\text{s}^{-1}$, e.g.][]{Hills:1980fx}, i.e.
\begin{equation}
	\tau \sim \frac{r_\text{pl,g}}{10\,\text{pc}/\text{Myr}} \,.
\end{equation}
We stress that the impact of the gas expulsion time scale depends on $\tau$ expressed in units of the cluster-forming region crossing time, $t_\text{cr}$:
\begin{equation}\label{eq:crossingtime}
  t_\text{cr} = \sqrt{\frac{8\,r_\text{vir}^3}{G M_\text{st}}}
\end{equation}
where the virial radius obeys
\begin{equation}
  r_\text{vir} = \frac{16}{3 \pi} r_\text{pl}
\end{equation}
in a Plummer model with a Plummer radius $r_\text{pl}$. 
Gas removal can be expected to be very rapid, and faster than the dynamical crossing time of the cluster-forming region \citep{Whitworth:1979uc}.
In the present calculations, $\tau$ takes the following values
\begin{equation}
\tau \in \left\{0.001\textbf{,}\ 0.15\textbf{,}\ 0.3\textbf{,}\ 0.6\textbf{,}\ 1.2\right\}\,t_\text{cr} \,.
\end{equation}
Table \ref{models_tab} gives $t_\text{cr}$ in Myr. 
The time-delay $t_\text{D}$ is adopted from \cite{2001MNRAS.321..699K} and corresponds to the gas-confinement time until the HII region erupts.
At a time $t = t_\text{D}  +\tau$, half of the initial gas mass is driven out.
In that respect, $\tau$ differs from the gas expulsion time scale adopted in \citet{2001ASPC..230..311G} and \citet{Parmentier:2008ew}, where it is defined as the time when gas removal is complete.

\subsection*{Alternative analytical model}
One alternative model representing the gas would be a Plummer model with constant mass and time-dependent Plummer radius $r_\text{pl,g}(t)$, e.g.
\begin{equation}
	r_\text{pl,g}(t) = \left\{\begin{array}{ll}
	r_\text{pl,g}^{(0)}, &t<t_\text{D} \\
	r_\text{pl,g}^{(0)} \left(2\left(1+\frac{t-t_\text{D}}{\tau}\right)  - 1 \right)^{1/2}, &t \geq t_\text{D}\,,
	\end{array}\right.
\end{equation} 
where $r_\text{pl,g}^{(0)}$ is the Plummer radius at $t<t_\text{D}$. This expression is derived from the condition to have, at the time $t$, the same gas mass left within a sphere about the centre of the system with radius $r_\text{pl,g}^{(0)} = r_\text{pl,g}(0)$ as in the first gas model.
\begin{figure}
	\includegraphics[width=0.47\textwidth] {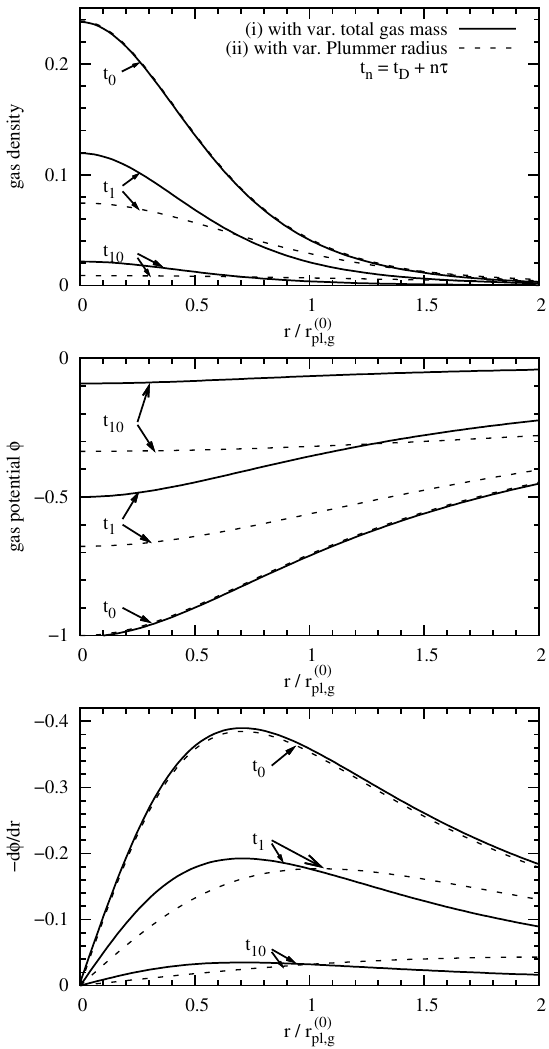}
	\caption{
		Comparison of two different analytical gas expulsion models. Both models assume that the initial gas distribution (at $t<t_\text{D}$) follows a Plummer model. The solid line corresponds to a model where the gas mass $M_\text{g}(t)$ is decreased with time (Eq. \ref{eq:gas-mass-vs-time}). In the alternative model (dotted line), the total gas mass is assumed to be constant, and the Plummer radius $r_\text{pl,g}$ is increased with time after $t>t_\text{D}$. The first panel shows the assumed gas density $\rho_\text{g}(r)$, while the second panel presents the resulting potential $\phi_\text{g}(r)$, and the third panel shows the resulting radial force, $m\ddot r = -\d\phi_\text{g}(r) / \d r$.
		The plots compare the gas models at different times, $t_{0}=t_\text{D}$, $t_{1}=t_\text{D}+\tau$, and $t_{10}=t_\text{D}+10\tau$.
	}
	\label{fig:gas-expulsion-models}
\end{figure}
In Fig.\,\ref{fig:gas-expulsion-models}, the corresponding density and potential are plotted for $t \in \left\{t_\text{D}, t_\text{D} + \tau, t_\text{D} + 10\tau\right\}$ (dashed line). 
Compared to the model based on Eq. \ref{eq:acceleration} we use in our calculations (solid line), the gas mass is not just reduced but the gas is driven outwards. 
As a consequence, in the model with variable Plummer radius the gas is removed faster from the central region (see Fig.\,\ref{fig:gas-expulsion-models}, first panel), and the potential remains deeper (second panel) so that cluster will disrupt more slowly as long as the Plummer radius of the gas is smaller than the tidal radius.
The maximum of the rejecting force, $-\d\phi(r)/\d r$, resulting from the gas potential can be found at
\begin{equation}
	r_\text{max}(t) = \frac{r_\text{pl,g}(t)}{\sqrt 2} \, .
\end{equation}
Thus, increasing the Plummer radius, $r_\text{pl,g}(t)$, means increasing the radius $r_\text{max}(t)$ at which the rejecting force is maximal, resulting in a larger, less dense core. 

In the present work, we use the decreasing mass model, because we assume the star cluster models to be initially not mass-segregated. In this case, it is more natural to remove the gas uniformly, following the distribution of the OB stars. \\


    \section{Initial conditions}
Five models with different initial properties (total mass and half-mass radius) are calculated.
Model III is based on model A of \citet{2001MNRAS.321..699K}.
The properties of all models are listed in Tab.\,\ref{models_tab}.

As for the tidal field, a Milky Way potential is adopted. The cluster has a distance to the Galactic centre of 8.5\,kpc and moves with a velocity of 220\,km\,s$^{-1}$ on a circular orbit through the Galactic disc. 

\subsection{The initial stellar population}
The stars are initially distributed in mass according to the canonical two-part power-law IMF \citep{2001MNRAS.322..231K}.
The distribution function $\xi(m)$ is given by 
\begin{equation}
\xi(m) \propto m^{-\alpha},
\end{equation}
where 
\begin{equation}
\alpha =\left\{\begin{array}{lrl}
			1.3,  & 0.08 \leq &m/\Msun < 0.5, \\
			2.3,  & 0.5 \leq &m/\Msun < 100 .
           \end{array}\right.
\end{equation}
$\xi(m)\d m$ is the number of stars with mass $m \in \left[m,m+\d m\right]$. 
For computational feasibility, there are no primordial binaries. This is a valid approximation, because they would not have an important effect on the surface brightness. 

\subsection{Cluster model}
Like the gas, the density distribution of the stellar population at $t < t_\text{D}$ with stellar mass $M_\text{st}$ is assumed to follow a Plummer model with a half mass radius, $R_{0.5}$, that is identical to that of the gas. 
The properties are listed in Tab.\,\ref{models_tab}. The models are initially not mass-segregated.

\renewcommand\arraystretch{1.5}
\begin{table}
  \caption{ 
  		Initial properties of the models.
		Note that model III here is model A of \citet{2001MNRAS.321..699K}.
  }
  \label{models_tab}
    \begin{tabular}{ cccccccc}
  \hline 
   &
  $M_\text{st}  \left[\Msun\right]$ & 
  $N$ & 
  $\left<m\right>  \left[\Msun\right]$ & 
  $R_{0.5} \left[\text{pc}\right]$ &
  $t_\text{cr} \left[\text{Myr}\right]$ \\
  \hline 
  I   	& 787	&	1343		&	0.59	&	0.263&	0.15	\\
  II   	& 787	&	1357		&	0.58	&	0.45	&	0.33	\\
  III 	& 3700 	& 	6382 	& 	0.58 & 	0.45 &	0.15	\\ 
  IV	& 18893	&	32578	&	0.58	&	0.45	&	0.067\\
  V 	& 18893	&	32122	&	0.59	&	0.77	&	0.15	\\ 
  \hline
  \end{tabular}

  \medskip
  This table gives an overview of the models we use. $M_\text{st}$ is the total stellar mass before gas expulsion onset, $N$ the number of stars, $\left<m\right>$ the mean stellar mass, $R_{0.5}$ the half-mass radius, and $t_\text{cr}$ the crossing time. The gas expulsion time scale $\tau$ (see Section \ref{gas_model}) takes values of 0.001\,$t_\text{cr}$, 0.15\,$t_\text{cr}$, 0.29\,$t_\text{cr}$, 0.59\,$t_\text{cr}$, and 1.18\,$t_\text{cr}$ in each model. 
\end{table}


    \section{Analysis steps}
For every time step, the Lagrange radii as well as the surface brightness are determined using the mass, position and total luminosity of the stellar ensemble. The density centre is calculated by the method of \cite{1985ApJ...298...80C}.
Mass loss due to stellar evolution is already taken into account by the simulation.

In order to analyse the time-evolution of the surface brightness, the average surface brightness within a constant radius (1, 3, 5, 10, and 15\,Myr) is plotted vs. time.

In order to compute the surface brightness of a finite ensemble of stars the stellar luminosity evolution as a function of the initial stellar mass is required. 
We use the spectral evolution code P{\sc egase} by \cite{fioc1997a,fioc1999a}. This code allows a user-defined multi-part power law IMF. In order to extract the evolution for a single star with mass $m_0$, we define a Dirac-IMF, $\xi(m) = \delta(m-m_0)$, and approximate it with a
single-part power law, $\xi(m)\propto m^{-\alpha}$, with slope $\alpha=0$ between $(1-0.001)m_0$ and $(1+0.001)m_0$ and zero otherwise. The masses $m_0$ are the same as the masses for which stellar evolution models are implemented into P{\sc egase}. 
A two-dimensional table (the stellar luminosity as a function of the initial stellar mass and the age of the star) for the passbands FUV, B, U, and V are constructed. A luminosity for an arbitrary initial stellar mass and age is obtained by two-dimensional linear interpolation of the luminosity table.


    \section{Results}
    As we analyze the results of the performed computations qualitatively, only model III is presented in detail, because the remaining models show qualitatively the same results. 
    
\begin{figure*}
	\includegraphics[width=\textwidth] {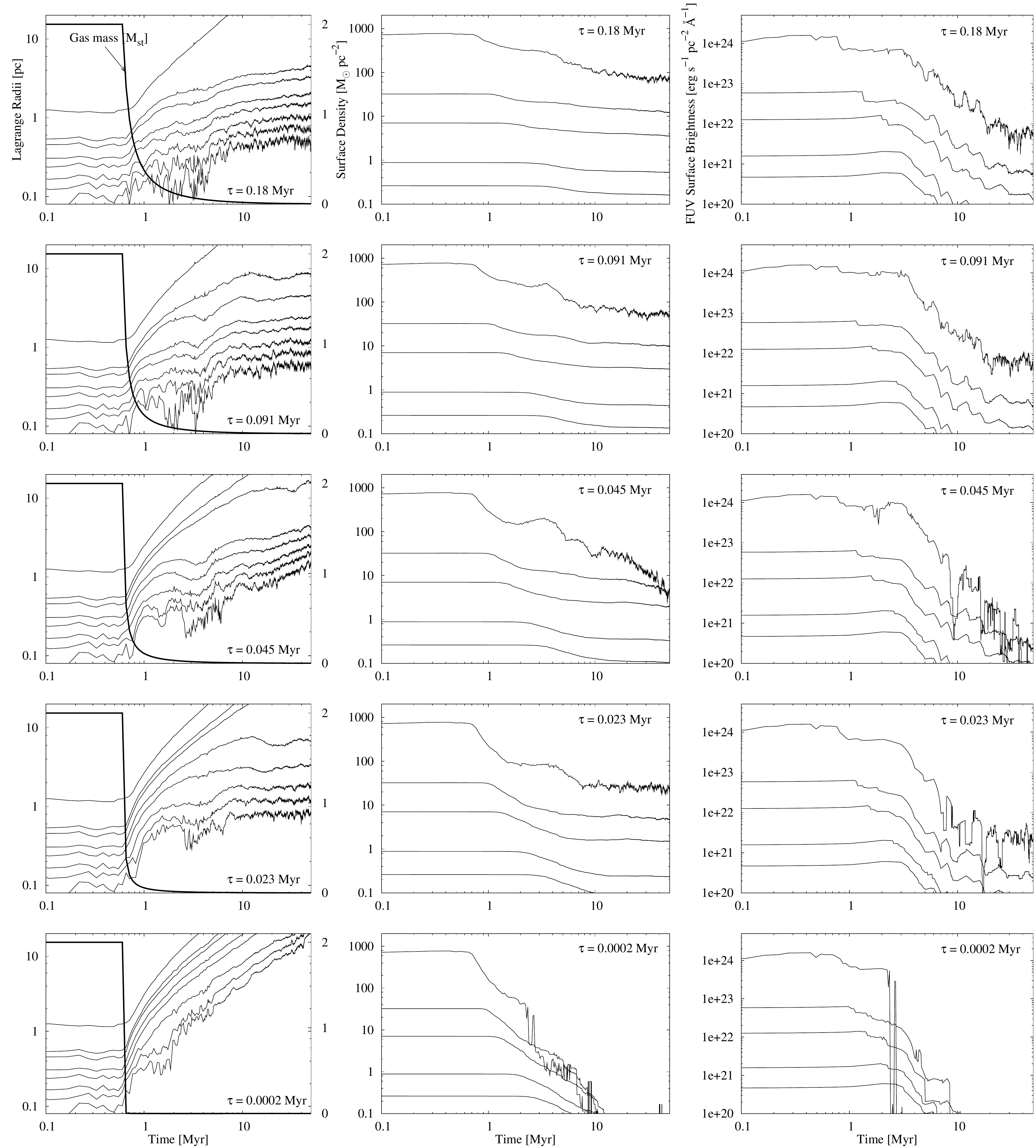}%
	\caption{\textbf{Model III}
		These plots show the main results of the simulations. The left column presents the Lagrange radii (within each panel bottom-up: 2\%, 5\%, 10\%, 20\%, 30\%, 40\%, 50\%, 80\%). From bottom to top, the gas expulsion rate slows: the smaller $\tau$, the quicker gas expulsion.
		The solid thick curve describes the mass of the residual gas, which is represented by the second $y$-axis and is given in units of the total stellar mass before gas expulsion onset.
		In the middle and right columns, the evolution of the surface mass density and the surface brightness within the inner 1, 3, 5, 10 and 15\,pc (solid curves, top to bottom) are presented. 
	}
	\label{results3}
\end{figure*}
\begin{figure*}
	\includegraphics[width=\textwidth] {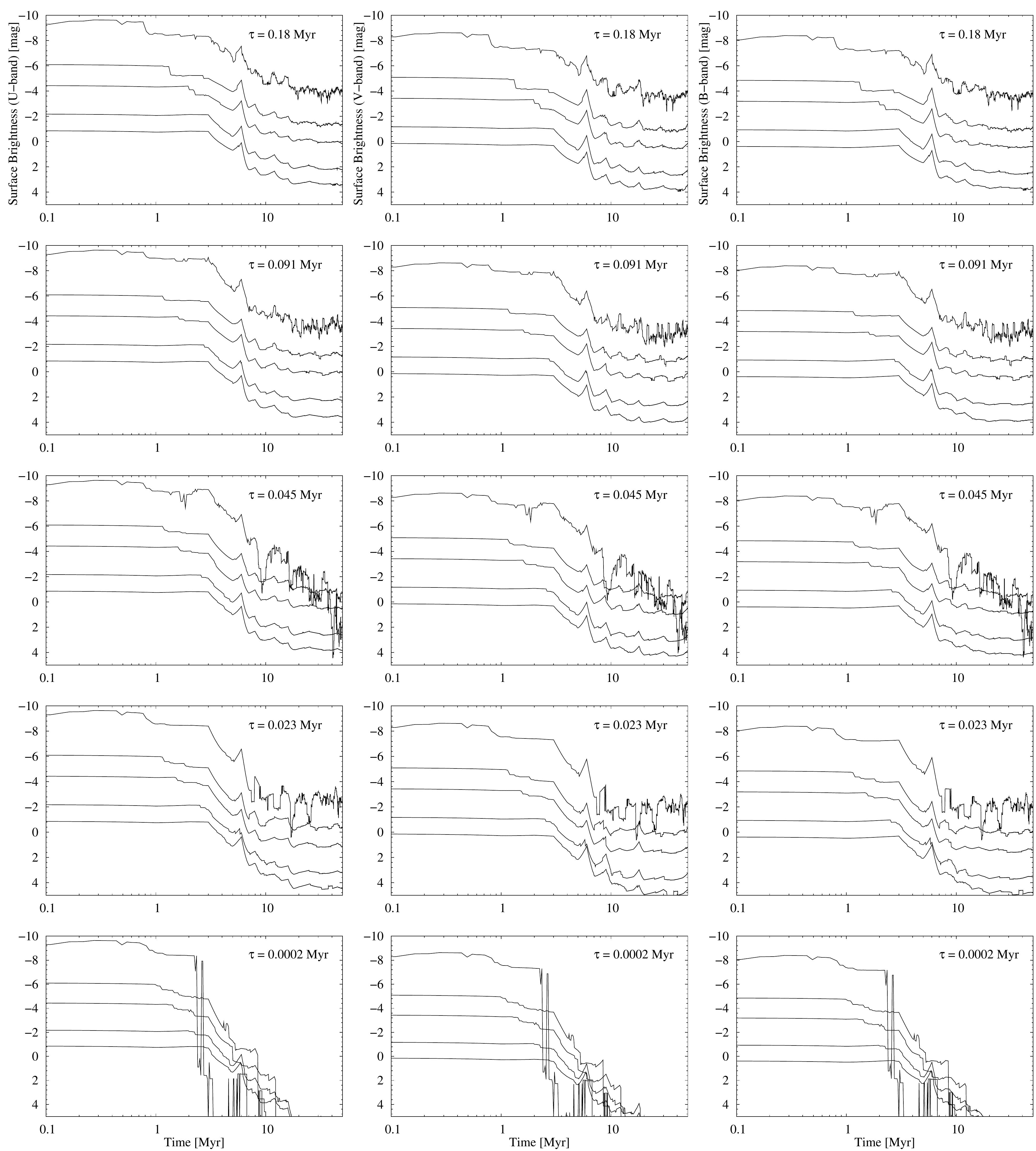}%
	\caption{\textbf{Model III} Complemental to Fig.\,\ref{results3}, this plot shows the surface brightness in the U, V, and B bands.}
	\label{results3uvb}
\end{figure*}
The computational results are shown in Figs.\,\ref{results3} and \ref{results3uvb}. As a check of consistency, the evolution of the Lagrange radii is compared between model A of \cite{2001MNRAS.321..699K} and model III here. They show the same characteristics: the expansion during and immediately after gas expulsion, followed by re-virialisation.

In model III, the Lagrange radii, the surface mass density, and the surface brightness in the four passbands specified above are plotted vs. time (six columns in total). 
The first column shows the Lagrange radii which document the expansion history of the cluster. Columns 2--6 contain the time-evolution of the average surface mass density/brightness within constant radii (1, 3, 5, 10, and 15\,pc). Each row presents a computation of model III with a different gas expulsion time scale $\tau$.

\begin{figure}
	\includegraphics[width=0.48\textwidth] {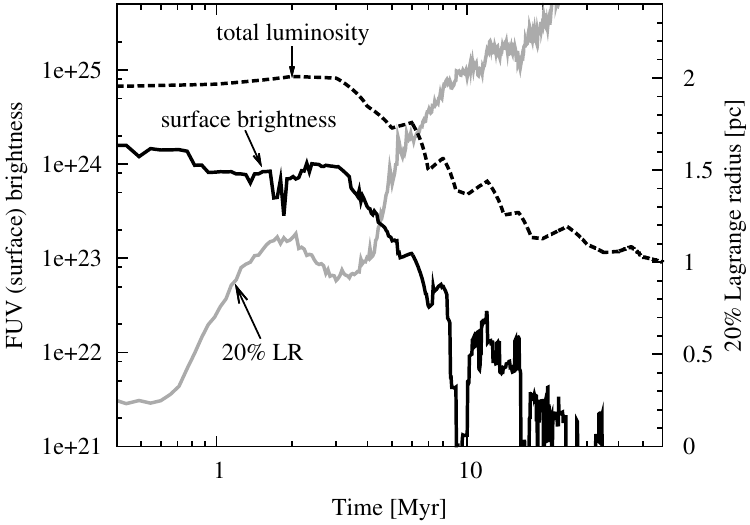}
	\caption{
		This plot illustrates the effects of stellar evolution and stellar dynamics on the cluster surface brightness. It shows the total FUV luminosity of all stars located within the innermost 1\,pc radius (dashed line, in erg\,$\text{s}^{-1} \text{\AA}^{-1}$) and the FUV surface brightness of a projected area with the same radius (solid line, in erg\,$\text{s}^{-1} \text{\AA}^{-1} \text{pc}^{-2}$) on the left $y$-axis. 
		In order to relate the surface brightness evolution to the re-virialisation process, the 20\% Lagrange radius (LR, grey line) is shown on the right $y$-axis.
	}
	\label{stellar_evolution}
\end{figure}
Figure \ref{stellar_evolution} shows the 20\% Langrage radius, the total cluster luminosity and the average FUV surface brightness 
within a projected 1\,pc radius circle around the density centre of the system.
On the basis of this figure, we want to discuss the interaction of cluster expansion and total stellar luminosity evolution and, accordingly, the dependencies of the surface brightness on stellar dynamics and stellar evolution.

\renewcommand\arraystretch{1.5}
\begin{table}
  \caption{ 
  		Re-increase of the surface brightness due to core re-virialisation in the V-band.
  }
  \label{tab:brightness_changes_revirialisation}
    \begin{tabular}{ccccc}
  \hline 
  Model & $\tau$\,[$t_\text{cr}$] & SB$_0$\,[mag] & SB$_\text{min}$\,[mag] & SB$_\text{max}$\,[mag] \\
  \hline 
	III	& 0.6 	& $-8.5$	& $-7.41$ 	& $-7.81$ \\ \hline
	IV	& 0.001 	& $-9.1	$	& $-8.3$ 	& $-8.7$ \\
	IV	& 1.2 	& $-9.1$	& $-8.5$ 	& $-8.8$ \\ \hline
	V	& 0.001 	& $-9.5$	& $-6.8$ 	& $-8.0$ \\
	V	& 0.3 	& $-9.5$	& $-8.4$ 	& $-8.7$ \\
	V	& 0.6 	& $-9.5$	& $-7.7$ 	& $-8.7$ \\
	V	& 1.2 	& $-9.5$	& $-7.8$ 	& $-8.5$ \\
  \hline
  \end{tabular}
  
  \medskip
  The table gives an overview of all calculations with a positive re-increase of the surface brightness within the inner 1\,pc due to core re-virialisation. Only 7 of 25 calculations show this behaviour. Here, $\tau$ is the gas expulsion time, SB$_{0}$ the initial surface brightness at $t<t_\text{D}$, SB$_\text{min}$ corresponds to the minimum surface brightness before re-virialisation at $t_\text{min}$, and SB$_\text{max}$ to the maximum surface brightness at $t_\text{max}$ after re-virialisation.
\end{table}
While the cluster starts to expand after the time $t_\text{D}$ has passed, the total luminosity of all stars increases clearly until $t\approx 2\,\text{Myr}$. It reaches its maximum near the core re-virialisation phase, depending on the model. From then on, the total stellar luminosity (dashed line) decreases distinctly as massive stars evolve.
Although the contraction of the core can be weakly detected as a bump in the surface mass density (see Fig.\,\ref{results3}, middle column), it is covered in the surface brightness by luminosity changes due to stellar evolution so that stellar dynamics play only a minor role in that context. Therefore, we can conclude that the re-virialisation in the core does not result in a significant re-increase of the central cluster surface brightness. 

To quantify the latter finding in all models (five models, each with five different values of $\tau$, i.e. 25 calculations in total), we do the following analyses. 
First, look at the surface mass density and detect the re-increase of the surface mass density within the inner 1\,pc which results from the core re-virialisation. If there is a noticeable increase, remember the times at which the surface mass density reaches the corresponding local minimum, t$_\text{min}$, and maximum, t$_\text{max}$, and determine the surface brightness changes (in the V-band) between $t_\text{min}$ and $t_\text{max}$.
Table \ref{tab:brightness_changes_revirialisation} shows the resulting surface brightness changes of all calculations in which such a re-increase is detectable. 7 of 25 models match the requirements.
In model I and II, not a single calculation matching these criteria was found.
In model III, $\tau = 0.6\,t_\text{cr}$ results in a weak re-increase of the surface brightness. Model IV has also two calculations with a weak increase and model V shows a visible effect in four calculations. However, compared to the later surface brightness decrease due to stellar evolution, the detected changes, $\text{SB}_\text{max} - \text{SB}_\text{min}$, are fairly small and it seems unlikely that they lead to the addressed effect of falling below and (re-)exceeding the detection limit.

Apart from our main question we notice some general dynamical aspects of gas expulsion when comparing simulations with the same initial conditions and different gas expulsion rates. The expansion of the 90\% Lagrange radius does not depend strongly on the gas expulsion time scale, although this parameter covers a range of three orders of magnitude.
In contrast, the expansion of the inner Lagrange radii is very sensitive to the gas expulsion time scale.
The quicker the gas expulsion, the faster the expansion of the central part of the cluster and accordingly of the inner Lagrange radii (first column in Fig.\,\ref{results3}).	
This is due to the effective crossing time, which gets longer with increasing radius and is shortest in the cluster core. 
Consequently, in the case of slow expulsion re-virialisation influences the surface mass density less strongly, because the cluster remains bound anyway for a long period. For short expulsion time scales on the other hand, the core expands fast and the core has only little chance to re-virialise. For the assumed SFE, one can observe re-virialisation clearly only for a range of gas expulsion times $\tau$, depending on the model. 

The surface mass density plots show statistical fluctuations in the central region due to the small number of stars (the noisy shape of the upper line of the second row of Fig.\,\ref{results3} after a few million years). 
The noisy shape in the surface brightness plots on the other hand is mainly caused by the evolution of individual stars.


    \section{Conclusions}
In the present work, we investigate the evolution of the surface brightness of star clusters under the assumption that the residual gas of the cluster-forming region is driven out by evolving OB stars, leading to a strong expansion of the total cluster. Five different models of star clusters with different sizes and masses are numerically evolved, every model with a SFE of $1/3$ and with five different gas expulsion time scales $\tau$. 
Thereby we compare the influences of stellar dynamics and stellar evolution on the surface brightness. 
Based on the expansion histories the question came up whether the impact of the re-virialisation process is sufficient to let the cluster core disappear under a given detection limit owing to its expansion, then re-appear after re-virialisation. From our computations we conclude that the surface brightness is not changed significantly by this process, because it is clearly dominated by stellar evolution.    

Independent by this finding, we can draw conclusions of the expansion behaviour. We find that the expansion of the inner parts of a star cluster is more sensitive to the gas expulsion time scale $\tau$ than the outer parts are. While the outer Lagrange radii are only weakly responsive to changing $\tau$, the inner Lagrange radii react very sensitively. E.g. the core expands moderately in the case of slow expulsion, while a quick expulsion on the other hand can lead to a rapid disruption of the core.


    \section{Acknowledgements}
    This work is a part of FL's Diploma thesis at the Argelander-Institut f\"ur Astronomie (Bonn, Germany).\\
    GP acknowledges support from the Humboldt Foundation and from the Max-Planck-Institut f\"ur Radioastronomie (Bonn, Germany) in the form of Research Fellowships.


\end{document}